\documentclass[11pt]{article}
\usepackage{geometry}                % See geometry.pdf to learn the layout options. There are lots.
\usepackage{graphicx}
\usepackage{amssymb}
\usepackage{epstopdf}
\usepackage{fixltx2e}
\usepackage{amsmath}
\usepackage{natbib}
\usepackage{hyperref}
\usepackage{rotating}
\hypersetup{urlcolor=blue, pdfborder=0 0 0}
\usepackage{setspace}
\usepackage{authblk}

\newcommand{\degree}{\ensuremath{^\circ}}

\title{Transient coupling relationships of the Holocene Australian monsoon}
\author[1]{F. H. McRobie\thanks{Corresponding author: fiona.mcrobie@research.uwa.edu.au}}
\author[2]{T. Stemler}
\author[1]{K.-H. Wyrwoll}
\affil[1]{School of Earth and Environment, University of Western Australia, Crawley, 6009, Australia}
\affil[2]{School of Mathematics and Statistics, University of Western Australia, Crawley, 6009, Australia}%\\ Submitted to Quaternary Science Reviews \\ January 2015}

\date{Accepted for publication in Quaternary Science Reviews \\ May 2015}                                           % Activate to display a given date or no date

\begin{document}
\bibliographystyle{apa}
%\linenumbers
\maketitle

\begin{abstract}
The northwest Australian summer monsoon owes a notable degree of its interannual variability to interactions with other regional monsoon systems. Therefore, changes in the nature of these relationships may contribute to variability in monsoon strength over longer time scales. Previous attempts to evaluate how proxy records from the Indonesian-Australian monsoon region correspond to other records from the Indian and East Asian monsoon regions, as well as to El Ni\~no-related proxy records, have been qualitiative, relying on `curve-fitting' methods. Here, we seek a quantitative approach for identifying coupling relationships between paleoclimate proxy records, employing statistical techniques to compute the interdependence of two paleoclimate time series. We verify the use of complex networks to identify coupling relationships between modern climate indices. This method is then extended to a set of paleoclimate proxy records from the Asian, Australasian and South American regions spanning the past 9,000 years. The resulting networks demonstrate the existence of coupling relationships between regional monsoon systems on millennial time scales, but also highlight the transient nature of teleconnections during this period. In the context of the northwest Australian summer monsoon, we recognise a shift in coupling relationships from strong interhemispheric links with East Asian and ITCZ-related proxy records in the mid-Holocene to significantly weaker coupling in the later Holocene. Although the identified links cannot explain the underlying physical processes leading to coupling between regional monsoon systems, this method provides a step towards understanding the role that changes in teleconnections play in millennial- to orbital-scale climate variability.
\end{abstract}

\section{Introduction}
The northwest Australian summer monsoon, and the related circulation over the Maritime Continent (i.e. the Indonesian-Australian summer monsoon -- IASM), is a critical feature of the global low latitude circulation. It provides a global heat source, and is the primary region of latent heat release associated with both the Southern Oscillation and the Madden-Julien Oscillation (MJO; \citealp{mcbride_indonesia_1998, hung_factors_2004}). Despite its importance, the Australian summer monsoon, occurring over the northwest Kimberley region of Australia, is relatively shallow, with sensible heating only observed below 750 hPa \citep{hung_factors_2004}.  Monsoon precipitation is relatively low, with annual November to April precipitation over northwestern Australia ranging from a mean of 1200mm (Kimberley Coastal Camp; \citealp{bureau_of_meteorology_climate_2014}) in the northwest, to 500mm at the south (Jubilee Downs, Broome; \citealp{bureau_of_meteorology_climate_2014}), over a distance of some 500km. Such a relatively weak monsoon system, located at the southern margins of the more general IASM regime, should be sensitive to changes in forcing mechanisms acting at both the global and regional scale, and over short and long time scales.

While a range of considerations come into play (e.g. \citealp{chang_northeasterly_1979, hung_factors_2004, wheeler_impacts_2009}), the dominant control on the Australian summer monsoon relates to the controlling role of the thermal land--sea contrast that manifests itself in the heat lows that develop during the summer months. IASM strength is also tied to the latitudinal position of the Intertropical Convergence Zone (ITCZ), separating equator-ward easterlies from poleward westerlies. The monsoon regime is characterised by summer rainfall associated with low-level westerlies that extend from the equator to around 15\degree S. The position of these westerlies is associated with the monsoon trough, representing a broad zone of strong convective activity with generally westerly inflow and characterised by the occurrence of monsoon depressions and tropical cyclones, defining the southern edge of the IASM region. With the progression of the seasons there is a northward displacement of the ITCZ, such that by the boreal summer it is located well to the north of the Maritime Continent, and is now associated with the East Asian summer monsoon \citep{chen_variation_2004}.

It is the onset of westerly flow which defines the Australian summer monsoon circulation, and `active' monsoon phases are linked to the MJO, resulting in strong convective activity and precipitation over the monsoon region \citep{hung_factors_2004, wheeler_impacts_2009}. Interhemispheric interactions between the IASM and the Northern Hemisphere are provided by cold surges emanating directly out of the East Asian winter monsoon, and leading to strong convective activity in the South China Sea and over the wider IASM region (Chang et al., 1979). It has also been suggested that these cold surges may also be directed into the Arabian Sea, enhancing MJO activity \citep{wang_observed_2012}, which provides a link with the Northern Hemisphere. These relationships make it clear that the present IASM is driven by an ensemble of regional and global scale climate controls (e.g. \citealp{chang_northeasterly_1979, meehl_annual_1987, hung_symmetry_2004, wang_observed_2012}).

When considered over longer time scales, additional drivers at both the global and regional scale need to be introduced. Milankovich insolation forcing of global monsoon systems has been long recognised (e.g. \citealp{clemens_forcing_1991, bowler_variations_2001, wang_millennial_2008}). Coupled ocean-atmospheric modelling studies have sought to explain the response of the northwest Australian monsoon to direct insolation forcing \citep{liu_coupled_2003, wyrwoll_sensitivity_2007, wyrwoll_orbital_2012}. These results suggest that although precession dominates changes in Northern Hemisphere monsoon strength, the Australian monsoon response is also significantly impacted by ocean temperature feedbacks \citep{liu_coupled_2003} and tilt forcing\citep{wyrwoll_sensitivity_2007}. \cite{liu_coupled_2003} suggest that the enhanced Australian monsoon at 11,000 years BP, contrary to reduced summer insolation, is due to a combination of sea surface temperature feedbacks and inflows from a strong East Asian winter monsoon.

The interconnected nature of these coupling relationships provides evidence for the `global monsoon' model as advocated in recent literature \citep{trenberth_global_2000, wang_global_2009, wang_global_2012, wang_global_2014}. This concept has been advanced to portray monsoon activity as a single body of tropical convection migrating about the equator according to seasonal heating, and tied closely to the positioning of the ITCZ \citep{wang_global_2009, wang_global_2014}. Over longer time scales, a coherent response of regional monsoons to Milankovich insolation forcing is noted by \cite{kutzbach_simulation_2008}. Using an accelerated transient simulation spanning 284,000 years, the authors display a positive response in regional monsoon systems to orbital forcing, with lead/lag relationships driven by local land and sea surface temperature feedbacks. As such, the global monsoon model has been extended to the paleoclimate context to describe this somewhat synchronous response to orbital forcing \citep{ziegler_precession_2010} as well as abrupt events such as the Heinrich Stadials \citep{cheng_global_2012}.

Here, we use complex network theory to analyse relationships between the northwest Australian summer monsoon, related monsoon systems and likely forcing climate states. We explore these relationships within the context of the `global monsoon', and through this we seek to separate global, interconnected relationships and drivers from more local controls. Using this approach, we attempt to establish the changing nature of the dynamical coupling relationships of the Australian summer monsoon over Holocene time scales.

\section{Methods}
Complex network theory offers a method for identifying coupling relationships and long-range teleconnections by connecting `similar' data sets. As such, it provides a suitable approach to assess interactions between monsoon systems within the context of the global monsoon \citep{donges_complex_2009}. By defining a measure of similarity between climate time series, climate networks have been shown to provide insight into dynamical interactions beyond the scope of traditional statistical analysis (e.g. \citealp{donges_complex_2009, donges_relationships_2013,van_der_mheen_interaction_2013, peron_correlations_2014}). Measures of similarity include linear cross-correlation, mutual information, and event synchronisation between extremes \citep{donges_complex_2009, rehfeld_similarity_2014}. Applying complex network methods to modern climate data is relatively straightforward, due to the availability of gridded datasets and high-density observation networks, but they also provide a powerful technique for analysing paleoclimate time series. This is demonstrated by \cite{rehfeld_late_2013} who developed a paleoclimate network of the Indian and East Asian summer monsoons covering the past 1,100 years, demonstrating distinct changes in network structure between the Medieval Warm Period, Little Ice Age and present day. The application of these techniques is facilitated by the development of a Matlab toolbox (\citealp{rehfeld_similarity_2014}; \url{http://tocsy.pik-potsdam.de/nest.php}). Here, we first construct a climate network using modern convective indices to demonstrate the veracity of complex network theory to identify dynamically-based coupling relationships between climate systems. We then develop a method for creating paleoclimate networks using a range of proxy records. The resulting paleoclimate networks identify linkages at the global and regional scale, and demonstrate the transient nature of coupling relationships of the northwest Australian monsoon region throughout the Holocene.

\subsection{Data}
\begin{figure}
\includegraphics[width=\textwidth]{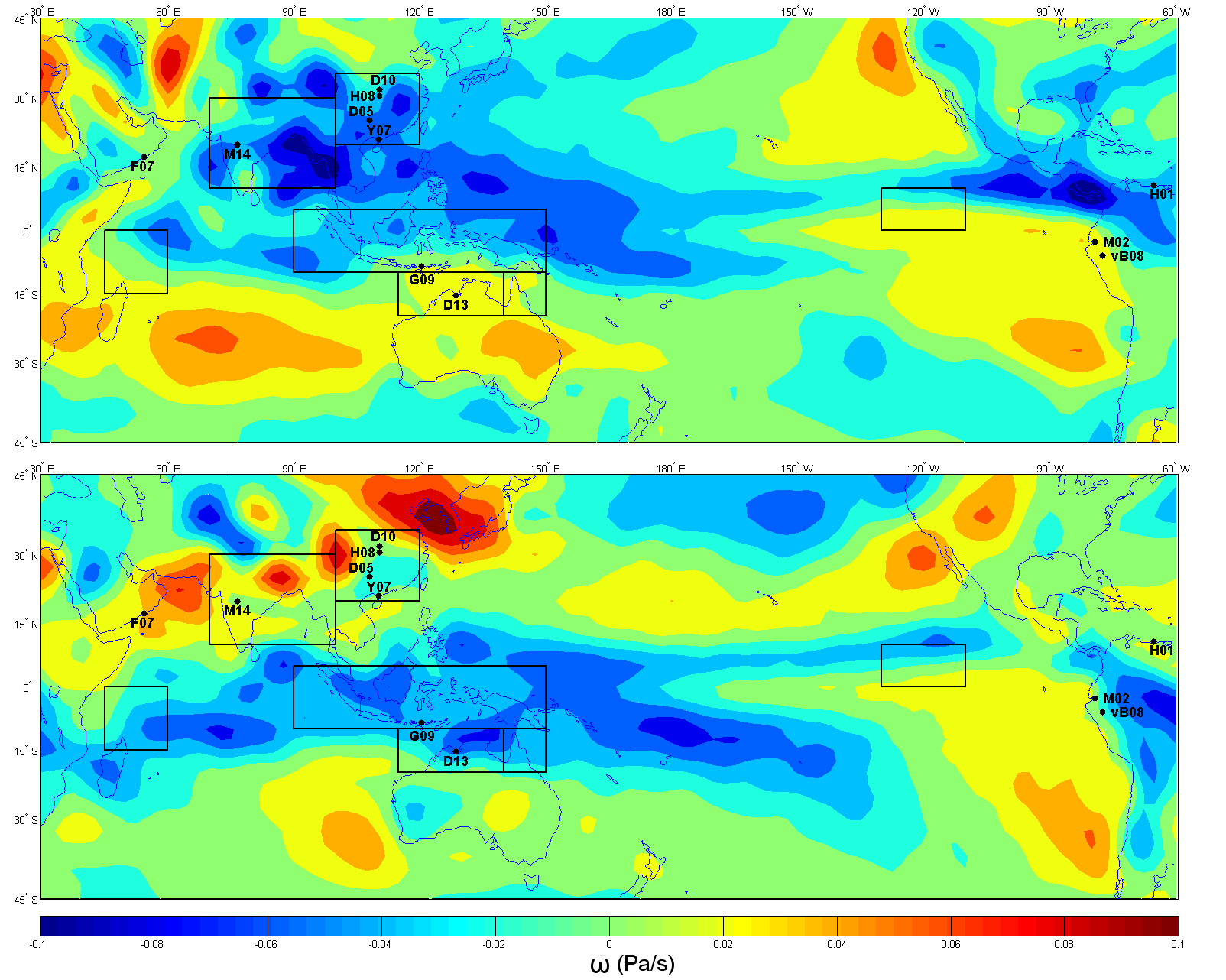}
\caption{a) DJF 1981--2010 500mb $\omega$ (NCEP Reanalysis, \citealp{kalnay_ncep/ncar_1996}). Also shown are the location of the boxes over which 500mb omega is averaged to produce modern convective indices (see Table~\ref{table:moderndata}), and the location of proxies; b) As above, for JJA.}
\label{fig:data}
\end{figure}

\begin{table}[!ht]
\caption{Modern climate data}
\centering
\begin{tabular}{ llcll }
\hline \hline
Code & Location & Lat/Lon Bounds & Season   \\
\hline
NWAus\textsubscript{DJF} & Northwest Australia & 10-20\degree S; 115-140\degree E & DJF \\
NEAus\textsubscript{DJF} & Northeast Australia & 10-20\degree S; 140-150\degree E & DJF \\
MC\textsubscript{DJF} & Maritime Continent & 5\degree N-10\degree S; 90-150\degree E & DJF \\
IO\textsubscript{DJF} & Western Indian Ocean & 0-15\degree S; 45-60\degree E & DJF \\
ISM\textsubscript{JJA} & Indian summer monsoon region & 5-25\degree N; 70-100\degree E & JJA \\
EASM\textsubscript{JJA} & East Asian summer monsoon region &10-20\degree N; 100-120\degree E & JJA \\
EEP\textsubscript{DJF} & East Equatorial Pacific & 0-10\degree N; 230-250\degree E & DJF \\
\end{tabular}
\label{table:moderndata}
\end{table}

\begin{table}[!ht]
\caption{Proxy records used in this analysis}
\centering
\resizebox{15.5cm}{!}{
\begin{tabular}{ llclll}
\hline \hline
Code & Location & Lat/Lon & Proxy Type & Reference & Average Time\\
& & & & & Step (years) \\
\hline
F07 & Qunf Cave, Oman & 17.17$\degree$N, 54.30$\degree$E & Speleothem $\delta^{18}$O & Fleitmann et al., 2007 &7.7\\
M14 & Lonar Lake, India & 19.98$\degree$N, 76.51$\degree$E & Multi-proxy  & Menzel et al., 2014 &18.8\\
H08 & Heshang Cave, China & 30.45$\degree$N, 100.42$\degree$E & Speleothem $\delta^{18}$O & Hu et al., 2008 &7.8\\
D05 & Dongge Cave, China & 25.28$\degree$N, 108.08$\degree$E & Speleothem $\delta^{18}$O & Dykoski et al., 2005 &14.7\\
Y07 & Lake Huguang Maar, China & 21.15$\degree$N, 110.28$\degree$E & Ti concentration & Yancheva et al., 2007 &0.8\\
& & & of lake sediment & &  \\
D10 & Sanbao Cave, China & 31.67$\degree$N, 110.43$\degree$E & Speleothem $\delta^{18}$O & Dong et al., 2010 &10.2\\
G09 & Liang Luar Cave, Indonesia& 8.52$\degree$S, 120.43$\degree$E & Speleothem $\delta^{18}$O & Griffiths et al., 2009 &10.1\\
D13 & Cave KNI-51, Australia & 15.30$\degree$S, 128.62$\degree$E & Speleothem $\delta^{18}$O & Denniston et al., 2013 &6.2\\
M02 & Laguna Pallacocha,  Ecuador& 2.77$\degree$S, 79.23$\degree$W & Red colour intensity & Moy et al., 2002 &0.8\\
& & & of lake sediment& &  \\
vB08 & Cueva del Tigre Perdido, Peru & 5.94$\degree$S, 77.31$\degree$W & Speleothem $\delta^{18}$O & van Breuklen et al., 2008 &19.4\\
H01 & Cariaco Basin & 10.70$\degree$N, 65.17$\degree$W & Ti concentration  & Haug et al., 2001 &5.6\\
& & &of marine sediment & &  \\
\end{tabular}
}
\label{table:proxies}
\end{table}

Our main aim is to capture coupling relationships of the Holocene Australian summer monsoon, but we first test the suitability of complex networks to identify dynamically-based coupling relationships using modern climate data. Seasonal convective indices are constructed using monthly values for 1948--2013 of mid-tropospheric (500mb) vertical velocity ($\omega$), a surrogate for convection (NCEP Reanalysis data provided by the NOAA/OAR/ESRL PSD, Boulder, Colorado, USA, from their web site at \url{http://www.esrl.noaa.gov/psd/}; \citealp{kalnay_ncep/ncar_1996}). In order to capture only coupling between deep convection, such as that associated with the monsoon circulation, we extract only three months of data from each year: December to February (DJF) or June to August (JJA), setting the values for the other nine months to zero (Table~\ref{table:moderndata}). This data is averaged over the regions covering northwest Australia (NWAus\textsubscript{DJF}), northeast Australia (NEAus\textsubscript{DJF}), the Maritime Continent (MC\textsubscript{DJF}), the western Indian Ocean (IO\textsubscript{DJF}), the Indian summer monsoon region (ISM\textsubscript{JJA}), the East Asian summer monsoon region (EASM\textsubscript{JJA}), and the Eastern Equatorial Pacific (EEP\textsubscript{DJF}). Note that the use of convective indices prevents the incorporation of the East Asian winter monsoon in our analysis. The East Asian winter monsoon is characterised by northerly winds driven by the Siberian High, causing cold surges outflowing over the South China Sea. There is some related convective activity in southern China, but insufficient to be captured by a convective-based index.

Following this, paleoclimate networks are produced for rolling 3,000 year windows at millennial intervals over the period 9,000 years BP to Present. We select proxy records (Table~\ref{table:proxies}) within the broad Indian Ocean-Pacific region according to high temporal resolution and low age uncertainty, as per \cite{rehfeld_similarity_2014}. Although one prefers a database comprised of a single proxy for reasons of comparability, one is often constrained by the number of proxy records available. We therefore combine speleothem \citep{fleitmann_holocene_2007, hu_quantification_2008, dykoski_high-resolution_2005, dong_high-resolution_2010, griffiths_increasing_2009, denniston_stalagmite_2013, van_breukelen_fossil_2008}, titanium \citep{yancheva_influence_2007, haug_southward_2001}, sediment \citep{moy_variability_2002} and multi-proxy \citep{menzel_linking_2014} data sets. The IASM region is represented in the proxy record database by two speleothem records, G09 (Liang Luar, Flores; \citealp{griffiths_increasing_2009}) and D13 (Cave KNI-51, northwest Australia; \citealp{denniston_stalagmite_2013}), both of which are interpreted as capturing monsoon precipitation trends and variation. The Chinese speleothem $\delta^{18}$O records (D05, H08, D10) have each been interpreted as a proxy for precipitation changes driven by the East Asian summer monsoon, while the Lake Huguang Maar record \citep{yancheva_influence_2007} has been discussed in the context of the East Asian winter monsoon and coupled to the IASM region in the modern climate through cold surges. We also include two widely used proxy records: the titanium concentration series from the Cariaco basin (H01; \citealp{haug_southward_2001}) has been cited in studies in the context of Holocene ITCZ positioning, and the Laguna Pallacocha sediment record from Peru (M02; \citealp{moy_variability_2002}) is a very widely used proxy for changes in El Ni\~no intensity and frequency over the last 12,000 years.

\subsection{Constructing complex networks}
\vspace{1em}
Estimating correlations between paleoclimate records is fraught with difficulty, and therefore an intuitive qualitative curve-fitting approach is typically employed. We apply methods widely accepted by statistical physicists which have been successfully applied in the context of financial markets \citep{zhuang_time_2014}, solar activity \citep{zou_long-term_2014}, disease dynamics \citep{zhang_hub_2010, wu_influence_2015, li_epidemic_2015}, and pigeon interactions in flight \citep{dieck_kattas_dynamical_2012, xu_reciprocal_2012}. In a climate or paleoclimate context, one may envisage such a network as a number of nodes, each corresponding to the site of a climate or paleoclimate data set. If a statistically significant `similarity' between two data sets is found, then an edge is drawn between the two nodes. More formally, for a database of $n$ time series, denoted $X_i$, we may describe the set of nodes as $V=\{v_{i}:i\in [n]\}$, and the set of edges is given by $E=\left\{e_{i,j}\right\}$ where $e_{i,j}=1$ is $X_i$ and $X_j$ are found to be statistically significantly `similar', and $e_{i,j}=0$ otherwise.We define similarity between two time series, $X_i$ and $X_j$, by mutual information, a nonlinear, symmetric (and thus non-directional) measure of how much information is shared between the two time series. Mutual information, $I(X_i,X_j)$ is given by:

$$I(X_i,X_j)=\sum_{x_i \in X_i} \sum_{x_j \in X_j} p(x_i,x_j) \log\left(\frac{p(x_i,x_j)}{p(x_i)p(x_j)}\right)$$ where $p(x_i)$ is the probability mass function of random variable $X_i$, and $p(x_i,x_j)$ is the joint probability mass function of $X_i$ and $X_j$. Note that, if $X_i$ and $X_j$ are independent, $p(x_i,x_j)=p(x_i)p(x_j)$, and hence mutual information is zero. If they are not independent, then the amount to which $p(x_i,x_j)$ differs from the product $p(x_i)p(x_j)$ provides a measure of the similarity of the two time series. We interpret this as a measure of coupling strength, with the information transfer between climate indices occurring through physical atmospheric flows and pressure-driven teleconnections. We choose mutual information over linear cross-correlation due to the nonlinear nature of the relationship between pairs of proxy records, visible in scatterplots. Cross-correlation can produce spurious results in this situation \citep{kantz_nonlinear_2003}. We note the bias inherent in the Gaussian mutual information estimate, as demonstrated by \cite{rehfeld_similarity_2014}. Irregular downsampling causes mutual information to be underestimated, and this bias increases rapidly when there are below 80--100 data points. Our analysis accounts for this: all the proxy records have more than 95 data points in any analysis window, except M14 (Lonar Lake, \citealp{menzel_linking_2014}) which has only 79 during the period 5,000--2,000 years BP and 86 during in the window 9,000--6,000 years BP. \cite{rehfeld_similarity_2014} suggest a bias-correction method for comparison with other similarity measures such as cross-correlation, but this is unnecessary here. Another similarity measure, the event synchronisation function, has been suggested to measure coupling between extreme events (\citealp{rehfeld_similarity_2014}). However, this requires the use of only the data beyond, say, the 90\% percentile, which for the modern convective indices would provide only around 20 data points in each time series. As a result, it is unsuitable for this analysis. Prior to estimation, the raw time series data are detrended using a Gaussian high-pass filter with a bandwidth equal to half the analysis window. For the paleoclimate time series, this is equal to 1,500 years, and means we remove frequencies slower than 1 oscillation every 9.4kyrs. Only the non-zero data points in the modern convective indices are detrended and used in the subsequent analysis. Since the teleconnections between the regional monsoon systems often involve a time delay of up to 12 months, in the modern data analysis, we estimate mutual information over a -12 to +12 month window, and take the maximum value.

Paleoclimate time series are often distributed along irregular time intervals due to sampling constraints. To account for this, a Gaussian kernel is used to `match' data in paired paleoclimate time series. \cite{rehfeld_comparison_2011} demonstrate that this reduces bias in the resulting mutual information estimate compared to linear interpolation. We use the Matlab toolbox of \cite{rehfeld_similarity_2014} to produce estimates of Gaussian kernel weighted mutual information, $I_{G}(X_i,X_j)$. This method does not produce symmetric estimates of $I_G$, but these asymmetric estimates do not imply directionality in the network, and are simply due to the unequal sampling rates of the two paleoclimate time series (Rehfeld et al., 2011). We therefore define:

$$I_{G}(X_i,X_j)=max(I_{G}(X_i,X_j),I_{G}(X_j,X_i))$$

We use a Monte Carlo approach to define statistically significant coupling relationships. For each modern or paleoclimate data set we generate a synthetic time series uncoupled to the others. Following \cite{rehfeld_late_2013}, we use an autoregressive model with one lag, Brownian motion with drift, to model the modern data sets and all but one of the paleoclimate data sets. The parameters -- linear drift and constant diffusion -- are estimated from the observed time series through linear regression. This time series is initially regularly spaced, and we downsample according to the time steps of the original, observed data set. The Laguna Pallacocha record from Ecuador (M02, \citealp{moy_variability_2002}) is not well suited to be modelled by Brownian motion. This time series is comprised of a number of large events which are registered well above a baseline level of near zero. We therefore introduce a Poisson process, to model the event time series defined by the 90\% quantile in the Laguna Pallacocha record.  This event time series is well approximated by a Poisson process ($\chi^{2}=1.85$, $p=10.12$, at a $95\%$ significance level).

The synthetic time series are used to determine statistically significant coupling relationships. Using the random time series models above, we generate synthetic time series corresponding to the observed data sets and create pairwise mutual information estimates. Repeating this 2000 times, we create 2000 randomly generated mutual information estimates for each pair of records. If the mutual information estimate calculated from the observed (modern or paleoclimate) data sets exceeds the 95th percentile of the randomly generated estimates, we claim there is a statistically significant coupling relationships between the two records. Only connections which are identified as statistically significant are displayed in the network.

Having constructed networks for the paleoclimate database (Table~\ref{table:proxies}) at 3,000 year windows throughout the last 9,000 years, we seek to evaluate changes in network density and structure. This may be attempted through a number of measures provided by graph theory (Newman, 2010). The \emph{degree}, $d_i$, of a node, $v_i$, describes the number of edges incident to the node, providing a description of how coupled the time series at $v_i$ is to other records in the network. Similarly, the \emph{network average degree}, $d_n$, is given by:

$$d_{n} = \frac{1}{n}\sum_{i=1}^{n}d_i$$ This quantifies the total amount of coupling within the network. In addition to total network connectivity, we consider the \emph{degree distribution}, the probability distribution of $d_i$ across the network. This allows us to determine whether the modern climate or paleoclimate records are all coupled to a similar degree, or whether a few records in particular are more dominant, potentially driving the broader monsoon network. In addition, we compare the observed degree distribution with the one from a random network where any two nodes are connected with probability $p=|E|/|E_{T}|$, where $E_T$ gives the total number of possible edges. Note that this produces a binomial degree distribution, taking as its parameters the number of nodes in the network and the number of connected edges as a fraction of the total number of possible edges. We then compare our observed degree distributions to this to identify any significant skew in the network connections.

\section{Testing complex networks using modern climate data}
\begin{figure}
\includegraphics[width=\textwidth]{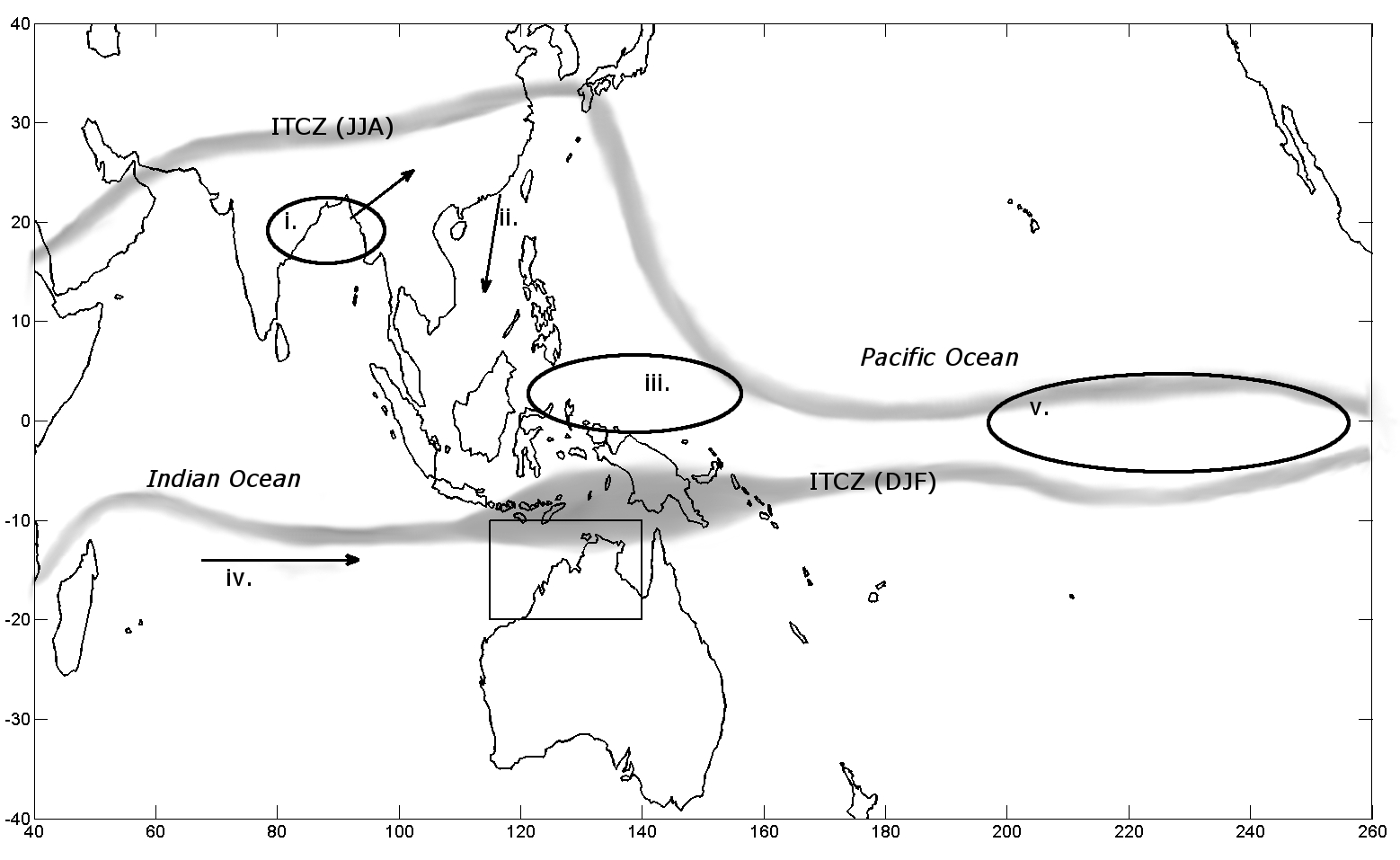}
\includegraphics[width=\textwidth]{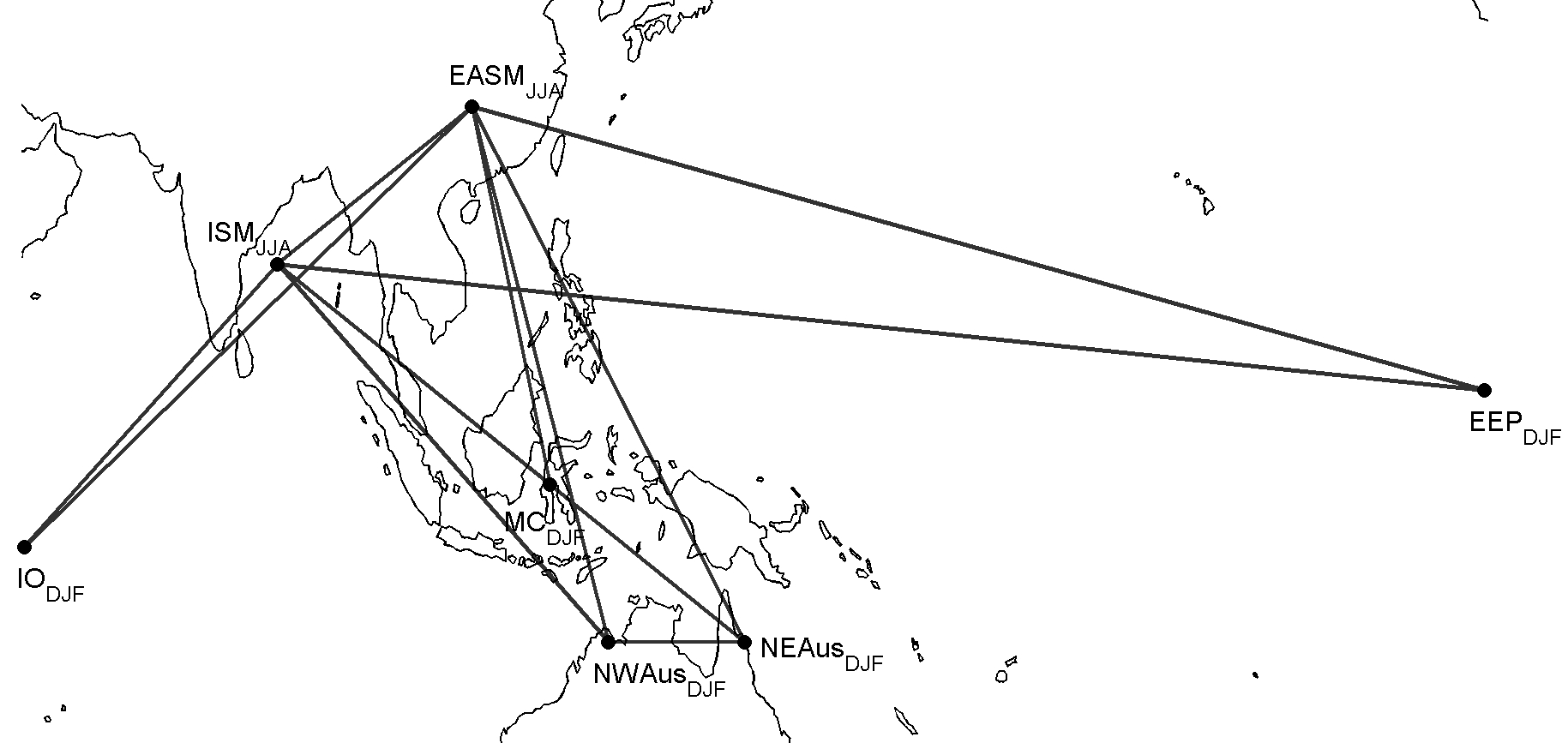}
\caption{a) Recognised interactions within the Asian-Australasian monsoon systems. We note the northward (JJA) and southward (DJF) positioning of the ITCZ, as well as i) the convective centre of the Bay of Bengal associated with the Indian summer monsoon, ii) northerly cold surges associated with the East Asian winter monsoon, iii) the Indo-Pacific Warm Pool, iv) the Ni\~no 3.4 region, and v) the Madden-Julian Oscillation. For reference, the box denotes the northwest Australian monsoon region as defined for this study. b) Modern climate network. Nodes are located in the centre of the zonal averaging region given in Table~\ref{table:moderndata}. Linked nodes are considered to be coupled at the 95\% significance level.}
\label{fig:modernMI}
\end{figure}

Because the coupling relationships between modern regional climate systems are relatively well known, they provide a suitable control against which we evaluate the use of complex networks. Figure~\ref{fig:modernMI}a displays the known dynamical mechanisms by which the Indian, East Asian and Indonesian-Australian monsoons interact. Note that, with the Indian and East Asian summer monsoons active in boreal summer (JJA), and the East Asian winter monsoon and Indonesia-Australian summer monsoon occurring in austral summer (DJF), some of these interactions occur with a seasonal lag.

In providing an explanation for the coupling relationships recognised in the `modern' data, we initially appeal to Chiang's (2009) framework for understanding the climate of the tropics. \cite{chiang_tropics_2009} outlines two models of tropical circulation. Providing a basic mechanism, the climate of the topics is explained by the Hadley circulation and its response to seasonal heating. Here, the migration of the ITCZ is primarily responsible for the distribution and timing of precipitation across the tropics. However, he notes that this simple explanation is only sufficient in an aquaplanet setting, and that the existence and location of land masses introduces regionality. It is within this second conceptual model that the regional monsoons are explained, asymmetries recognised, and the role of ENSO incorporated.

Figure~\ref{fig:modernMI}a depicts the ITCZ positioning in boreal and austral summer, with convergence-driven convective activity located over India, southern China and northern South America in JJA, while in DJF the ITCZ sits south of the equator, bringing convective activity to the Indonesian-Australian region and the Indian Ocean stretching from Indonesia to the north tip of Madagascar. Interactions between different regional features are also depicted. Outflows from the Bay of Bengal, associated with the Indian summer monsoon, are a key moisture source in the East Asian summer monsoon \citep{yihui_east_2005}. The Bay of Bengal is also the point of origin of convective centres which are displaced southwards over a number of months towards the Indonesian-Australian monsoon region \citep{meehl_annual_1987}. \cite{hung_symmetry_2004} demonstrate a correlation between the Indian and Australian monsoon regions, but describe a ``communication gap" between the two systems, with heavy (weak) Indian summer monsoon precipitation followed by heavy (weak) Australian monsoon precipitation. It is worth noting that this analysis combined the northwest and northeast of Australia into a single region, creating correlations with the El Ni\~no-Southern Oscillation (ENSO) which are likely due to the impact of ENSO on tropical northeast Australia. The relationship between ENSO and the Indian \citep{krishnamurthy_indian_2000} and East Asian \citep{wang_pacificeast_2000} summer monsoons is well established. Shifts in both the Walker and Hadley cells caused by warm (cool) sea surface temperature anomalies in the East Equatorial Pacific act to dampen (strengthen) the Indian summer monsoon, while El Ni\~no events establish Rossby waves travelling towards China, setting up a region of anticyclonic circulation over the Philippine Sea and suppressing East Asian summer monsoon convection. Finally, the East Asian winter monsoon establishes northerly winds over China, producing irregular low-level surges of cool air which travel southwards across the South China Sea and into the Indonesian sector \citep{chang_northeasterly_1979}. These surges are able to enhance convective activity, uplifting the already warm, moist air. A second outflow of cold air has been proposed to travel westward, flowing to the north of the Tibetan Plateau before being deflected southwards and across the Arabian Sea. \cite{wang_observed_2012} argue that this influx of cool air excites the MJO, thus acting as a secondary forcing mechanism on the Indonesian-Australian monsoon regime.

The mutual information-based climate network (Figure~\ref{fig:modernMI}b) captures many of the interactions outlined above. Due to the fact that the convective indices incorporate only boreal summer data from the East Asian region, we cannot capture any of the interactions involving the East Asian winter monsoon region. As noted in the methods section, this is intentional, as our $\omega$-based indices cannot capture East Asian winter monsoon strength. The climate network, however, does miss two accepted coupling relationships. We expect a link to be observed between the east equatorial Pacific (EEP\textsubscript{DJF}) and northeast Australia (NEAus\textsubscript{DJF}). There are two possible reasons for not capturing this in our analysis: either the link is missing because the signal cannot be distinguished through the noise of the climate time series, or the box size of the northeast Australian region (cf Figure~\ref{fig:data}) might be unable to accurately capture regional precipitation using the NCEP/NCAR reanalysis. This is an issue which translates into the paleoclimate context directly, as many proxy records will not provide a `pure' signal of monsoon-related precipitation, but rather capture a number of other climate and environmental changes.

The second link missing in the modern climate network is between the Maritime Continent (MC\textsubscript{DJF}) and northwest Australia (NWAus\textsubscript{DJF}). Given that the Indonesian-Australian summer monsoon extends across both regions, a coupling relationship between the two time series would be expected. However, \cite{haylock_spatial_2001} examine summer rainfall measured at 63 stations across Indonesia, demonstrating limited spatial coherence across the region, with no single forcing mechanism or predictor of wet season precipitation. As such, averaging DJF convective activity across the region is unlikely to produce an index which can be interpreted easily in the context of coupling relationships with other convective indices.

Despite the two missing network links, we have confidence in the ability of complex networks to capture coupling relationships between climate signals. The fact that there are no coupling relationships identified which do not correspond to any understood dynamical mechanisms further supports this, and validates the decision to produce networks based on mutual information. The spurious values which can arise when estimating cross-correlation between non-linear time series could lead to coupling relationships being identified which have no physical basis.

\section{Coupling relationships of the Australian summer monsoon over the last 9,000 years}
\begin{figure}
\centering
\includegraphics[width=0.8\textwidth]{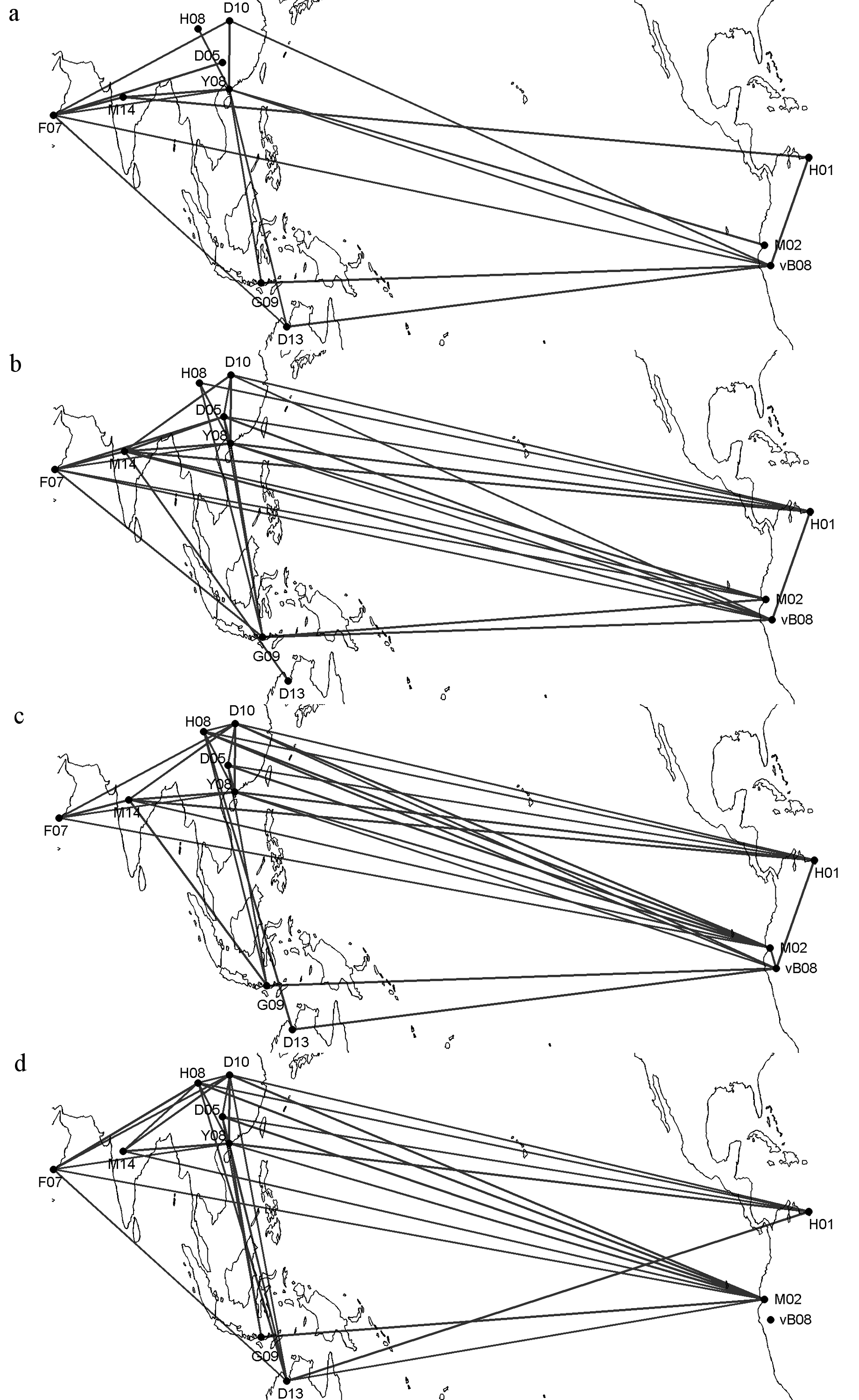}
\caption{Paleoclimate networks for \emph{a}: 9,000-6,000 yrs BP, \emph{b}: 8,000-5,000 yrs BP, \emph{c}: 7,000-4,000 yrs BP, \emph{d}: 6,000-3,000 yrs BP. Linked nodes are coupled at the 95\% level.}
\label{fig:I_G}
\end{figure}

\renewcommand{\thefigure}{\arabic{figure} (Cont.)}
\addtocounter{figure}{-1}
\begin{figure}
\centering
\includegraphics[width=0.8\textwidth]{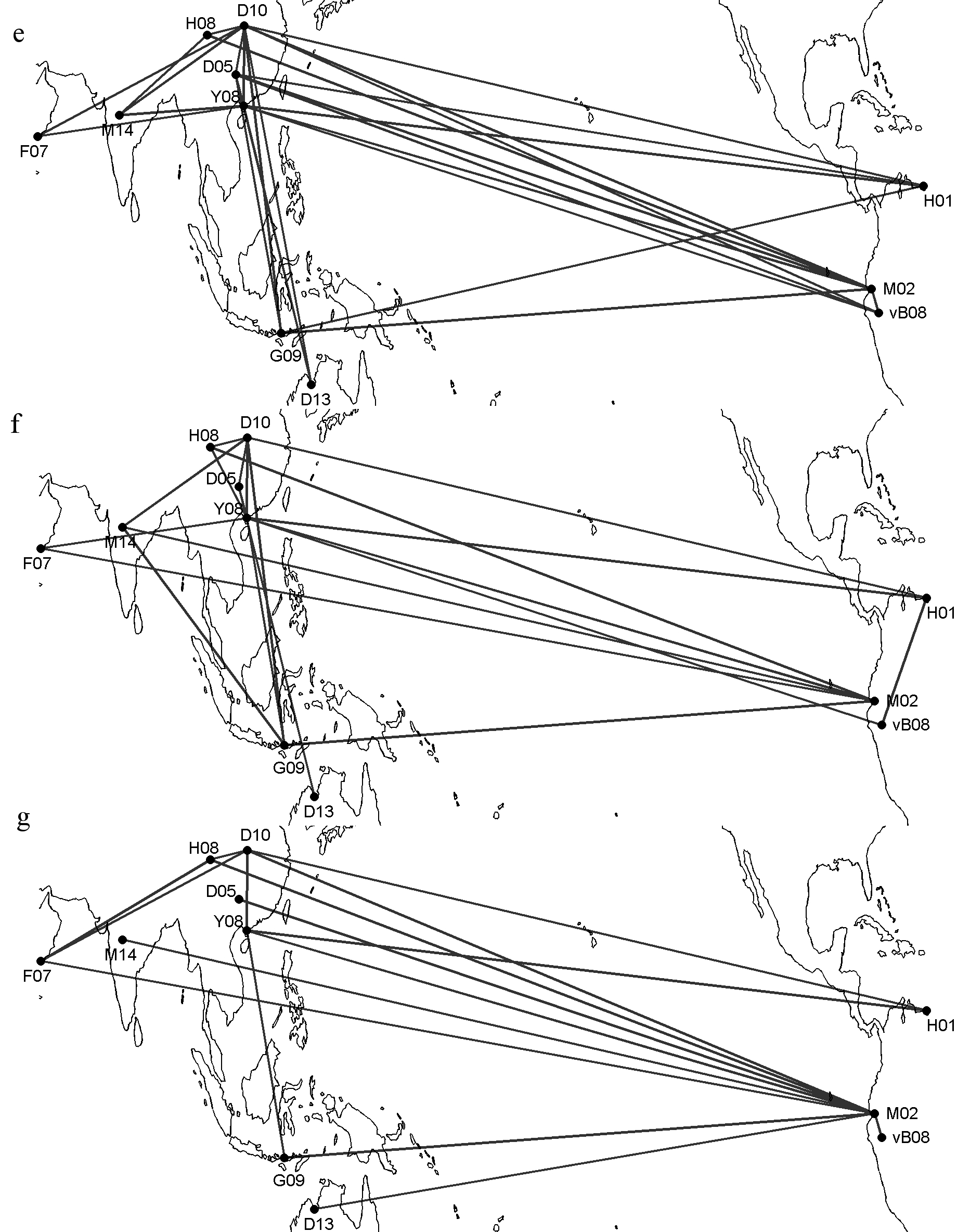}
\caption{\emph{e}: 5,000-2,000 yrs BP, \emph{f}: 4,000-1,000 yrs BP, \emph{g}: 3,000-0 yrs BP.}
\end{figure}
\renewcommand{\thefigure}{\arabic{figure}}

The Holocene combines, among other things, a period of changing solar insolation forcing, sea surface temperature feedbacks, and changes in land extent following the last deglaciation (cf. \citealp{lambeck_late_1990, liu_coupled_2003, jansen_early_2008}). We would therefore expect that complex networks may provide a formal demonstration of changing teleconnections throughout the Holocene. We demonstrated in section 3 that complex networks can capture coupling relationships between modern climate signals. We now present paleoclimate networks at 3,000 year windows, every millennium, from 9,000 yrs BP to Present. The network is populated by the database of proxy records from the Asian--Pacific region (Table~\ref{table:proxies}). Each edge in the network (Figure~\ref{fig:I_G}) identifies a statistically significant amount of information shared between two paleoclimate proxy records. We interpret these edges as dynamically-based coupling relationships between regional climate systems. With network edges identifying significant coupling relationships, the total degree of connectivity within the network should provide insight into the collective behaviour of regional monsoon systems during the Holocene. However, as our focus is on the Australian summer monsoon, we emphasise nodes G09 and D13 (corresponding to Liang Luar, Flores, Indonesia; \citealp{griffiths_increasing_2009}, and Cave KNI-51, northwest Australia; \citealp{denniston_stalagmite_2013}) to enable us to draw inferences regarding the coupling relationships of the Indonesian-Australian monsoon regime.

\subsection{Overall trends in network relationships}
\begin{figure}
\includegraphics[width=\textwidth]{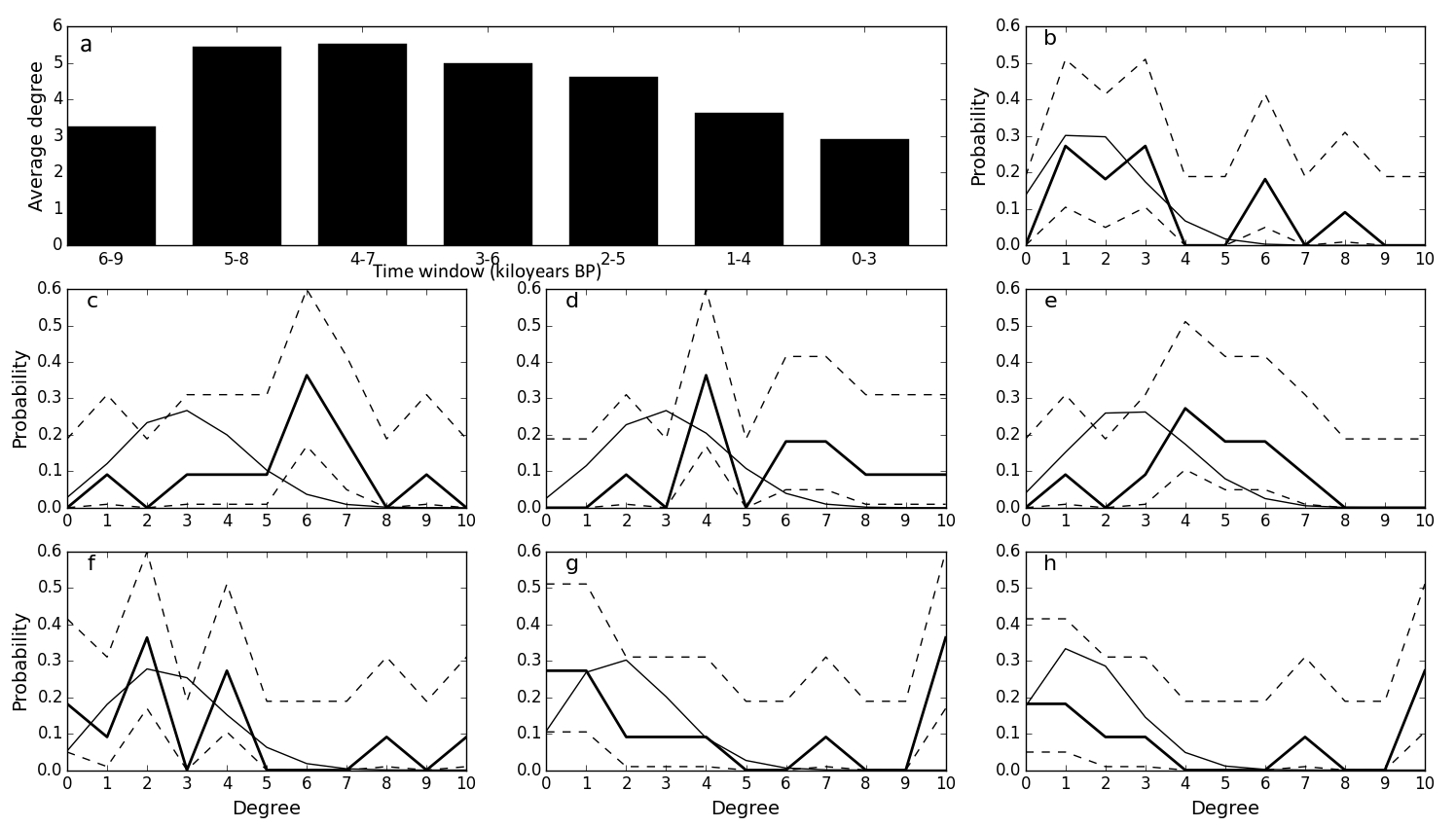}
\caption{a) Network average degree; b-h) Degree distributions for the observed paleoclimate network (black), with 90\% upper and lower confidence intervals (dotted lines), and that of a random graph with the same number of edges (grey).}
\label{fig:degree}
\end{figure}

The paleoclimate networks are observed to grow increasingly connected from 9,000--6,000 yrs BP to 6,000--3,000 yrs BP, and then decline steadily into the latest Holocene (Figure~\ref{fig:I_G}).  In particular, the nodes corresponding to the records H01 (Cariaco Basin; \citealp{haug_southward_2001}), M02 (Laguna Pallacocha, Ecuador; \citealp{moy_variability_2002}), vB08 (Cueva del Tigre Perdido, Peru; \citealp{van_breukelen_fossil_2008}), and Y07 (Lake Huguang Maar, China; \citealp{yancheva_influence_2007}) display high degree throughout the mid-Holocene, while only M02 is seen to be highly coupled in the later Holocene. We quantify total network connectivity using the average network degree. This highlights a trend of increasing degree until 6,000--3,000 yrs BP, followed by a return to nodes of lower degree (Figure~\ref{fig:degree}a). To determine if the coupling relationships are evenly distributed across the network we compare the observed degree distribution for each 3,000 year window with the one of a random graph. This random graph has the same number of nodes and edges as the observed network and a binomial degree distribution. The small dataset accounts for the large confidence error bands, which mean that we cannot state much, with confidence, about the difference between our observed networks and a random graph. However, during the mid-Holocene the degree distributions resemble that of a random graph which sits within the 90\% confidence bands (Figures~\ref{fig:degree}c -- e). This means that although the network has a higher degree during this period, coupling relationships are spread somewhat evenly across the region -- rather a situation where one proxy record dominates the network, instead there is a relatively equal amount of coherence between the regional monsoon systems. In the late Holocene (Figures~\ref{fig:degree}f -- h) there is increasing skewness, with a statistically significant deviation for degree greater than 7 from the distribution from a random graph. This demonstrates that although most proxy records have few coupling relationships with other records, a small number of records `dominate' the network, playing a more critical role. The Laguna Pallacocha (M02) record stands out here as being very highly connected, with coupling relationships identified with nearly every other record in the network.

\subsection{Coupling relationships of the IASM region}
The proxy records within the Indonesian-Australian monsoon region are G09 and D13, speleothem $\delta^{18}$O records located in Indonesia and northwest Australia respectively \citep{griffiths_increasing_2009, denniston_stalagmite_2013}. Both records have been interpreted as representing local precipitation, and tied directly to the IASM circulation. Despite this, no coupling relationship between the two records is identified in the paleoclimate networks (Figure~\ref{fig:I_G}). Possible explanations for this include the rising sea levels experienced during the deglaciation and into the mid-Holocene, and by the modern-day climatology which displays large spatial variations in monsoon precipitation across the region. To investigate this, we draw upon other proxy records from the region, unsuitable for our network analysis, but able to assist in a qualitative interpretation.

\begin{figure}
\includegraphics[width=\textwidth]{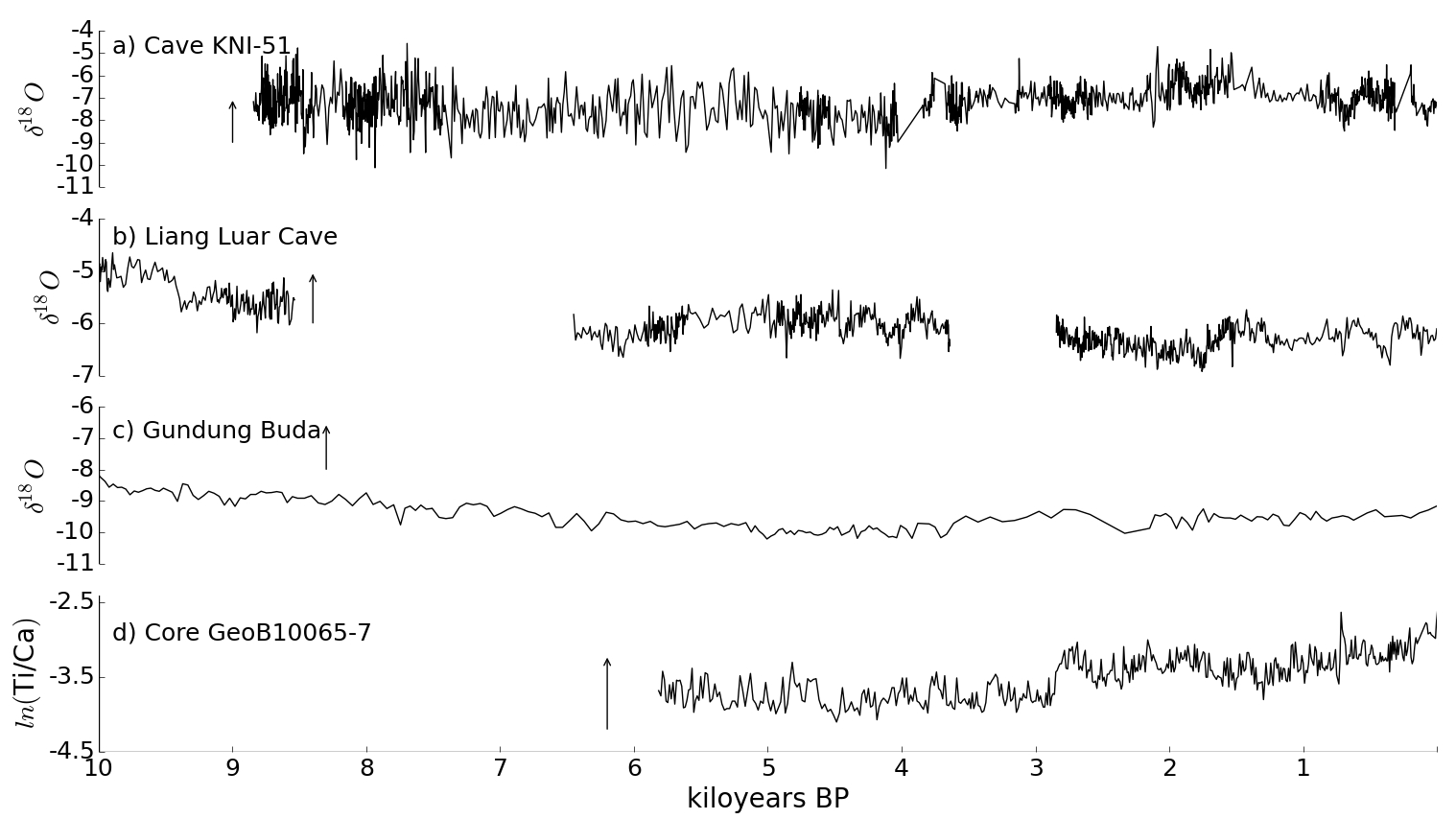}
\caption{Holocene proxy records from the Indonesian-Australian monsoon region. a) speleothem $\delta^{18}$O record from Cave KNI-51, northwest Australia  (D13; \citealp{denniston_stalagmite_2013}), b) speleothem $\delta^{18}$O record from Flores, Indonesia  (G09; \citealp{griffiths_increasing_2009}), c) speleothem $\delta^{18}$O record from Gunung Buda, Borneo \citep{partin_millennial-scale_2007}, d) sediment record from offshore Sumba Island, Indonesia \citep{steinke_mid-_2014}. The arrows denote in the direction of enhanced monsoon strength in the proxy record interpretation.}
\label{fig:IASMproxies}
\end{figure}

With the progressive increase in sea level from c. -130 m over the 20 000 years (summary in \citealp{murray-wallace_quaternary_2014}), land extent over the IASM region was significantly modified \citep{voris_maps_2000}. During our period of interest, records suggest a sea level of around 20m below PMSL at around 9,000 years BP, increasing to present day levels by about 7,000 years BP \citep{lewis_post-glacial_2013, murray-wallace_quaternary_2014}. Paleoclimate proxy studies from the IASM region suggest a broad-scale response to sea level changes in the early to mid-Holocene (Figures~\ref{fig:IASMproxies}a, b, c). \cite{griffiths_increasing_2009} interpret the strengthening precipitation signal from around 11,000 to 6,000 years BP in the Liang Luar (G10) speleothem record as evidence of a monsoon response to the flooding of the Sunda shelf. This is also noted by \cite{denniston_stalagmite_2013} in the Cave KNI-51 record (D13), with strengthening from 9,000 to 7,000 years BP possibly tied to Sahul and Sunda shelf flooding. A similar strengthening of the monsoon is also recognised in other speleothem records: Gunung Buda, north Borneo (Figure~\ref{fig:IASMproxies}c, \citealp{partin_millennial-scale_2007}) and Ball Gown Cave, northwest Australia \citep{denniston_last_2013}. These trends represent a multi-millennial scale response to sea level change, with monsoon strengthening observed over some 5,000 years.

The Indonesian (G09) and northwest Australian (D13) records remain uncoupled in the later networks, which clearly cannot be attributed to sea level rise. Instead, we look to the spatial variations observed in the modern IASM circulation \citep{haylock_spatial_2001}. Heterogeneity in monsoon-related precipitation across the IASM region has been attributed to possible local relief or island controls (e.g. \citealp{moron_spatial_2009}) as well as varying interactions with the Indo-Pacific Warm Pool and sensitivity to interannual ITCZ positioning (e.g. \citealp{partin_millennial-scale_2007}). Proxy records available in the late Holocene -- Flores, Indonesia (G09, \citealp{griffiths_increasing_2009}), Cave KNI-51, northwest Australia (D13, \citealp{denniston_stalagmite_2013}), Gunung Buda, northern Borneo \citep{partin_millennial-scale_2007}, and Lombok Basin, near Sumba, Indonesia \citep{steinke_mid-_2014} -- indicate no coherence in monsoon strength at millennial to centennial time scales (Figure~\ref{fig:IASMproxies}). \cite{denniston_stalagmite_2013} consider their Cave KNI-51 record to be anti-phased with the Flores record in the late Holocene, and note that the weakening of the monsoon observed in the northwest Australian record from 4,000 to 1,500 years BP is much less visible in the Indonesian record. The Lombok Basin record \citep{steinke_mid-_2014} has been interpreted as displaying a very different chronology, with an abrupt increase in precipitation at around 2,800 years BP, with this stronger monsoon regime continuing until 1,700 years BP. An explanation for the difference in sub-millennial scale events in the northwest Australian and Indonesian records during the late Holocene was considered by \cite{steinke_mid-_2014}. They suggested that modelling results of the solar minimum at around 2,800 years BP demonstrated varied responses across the IASM region, with strengthening across Indonesia, weakening in northern Australia, and a somewhat neutral response in Borneo. It is therefore reasonable that short-term fluctuations in the IASM paleoclimate proxy records might be uncoupled due to spatially heterogeneous responses to external forcing across the region, just as can be observed in the present day climatology.

Despite the differences between the Liang Luar (G09; \citealp{griffiths_increasing_2009}) and Cave KNI-51 (D13; \citealp{denniston_stalagmite_2013}) records, they display similar coupling relationships with other, non-IASM proxies (Figure~\ref{fig:I_G}). From 9,000--6,000 yrs BP to 4,000--1,000 yrs BP, these two records are coupled (with some inconsistency over time) to number of records forming a belt from the Arabian Pennisula to South America: Qunf Cave, Oman (F07; \citealp{fleitmann_holocene_2007}), Lonar Lake, India (M14; \citealp{menzel_linking_2014}), Lake Huguang Maar, southern China (Y07; \citealp{yancheva_influence_2007}), Cueva del Perdido, Peru (vB08; \citealp{van_breukelen_fossil_2008}), and Cariaco Basin (H01; \citealp{haug_southward_2001}). There is, additionally, some coupling between records in the Indonesian-Australian and East Asian summer monsoon regions during 6,000--3,000 yrs BP and 5,000--2,000 yrs BP. However, in the latest Holocene (3,000--0 yrs BP), the only record which both Liang Luar (G09; \citealp{griffiths_increasing_2009}) and Cave KNI-51 (D13; \citealp{denniston_stalagmite_2013}) are connected to is the Laguna Pallacocha record from Ecuador (M02; \citealp{moy_variability_2002}).

\subsection{Interpretation of IASM coupling relationships of the last 9,000 years}
The paleoclimate proxy records from Oman, India, southern China, Peru and Cariaco Basin have each been interpreted in the context of ITCZ positioning. The titanium concentration time series from Cariaco basin (H01; \citealp{haug_southward_2001}) and the speleothem oxygen isotope composition in Cueva del Tigre Perdido, Peru (vB08; \citealp{van_breukelen_fossil_2008}) are dependent on the position of the ITCZ during boreal winter. The Lake Huguang Maar sediment record in southern China (Y07) is interpreted as a proxy for the East Asian winter monsoon strength, determined over millennial scales by the position of the ITCZ \citep{yancheva_influence_2007}. The bioclastic climate index from Lonar Lake, central India (M14) is not explicitly interpreted as an indicator of ITCZ positioning, but the authors tie early Holocene Indian summer monsoon strength to ITCZ migration \citep{menzel_linking_2014, prasad_prolonged_2014}. Finally, Qunf cave in southern Oman (F07; \citealp{fleitmann_holocene_2007}) is situated at the southern edge of the modern position of the ITCZ in boreal summer. A northwards shift of the ITCZ causes southwesterly flow associated with the Indian summer monsoon to extend over the lower tip of the Arabian Peninsula, lifting the local temperature inversion and triggering deep convective precipitation. As the ITCZ shifts south, however, its role in connecting these two regions ceases. The coupling relationships identified between F09 and M14 from 9,000 to 4,000 years BP (Figure~\ref{fig:I_G}a-c) therefore support a northerly positioning of the ITCZ, as expected under Northern Hemisphere precessional bias. At the same time, the monsoon region of Indonesia and northwest Australia experiences a weak but strengthening monsoon, and the density of coupling relationships with `ITCZ-proxies' and the East Asian summer monsoon proxies suggests an Indonesian-Australian monsoon modulated by global-scale forcing.

The northwest Australian proxy from Cave KNI-51 (D13; \citealp{denniston_stalagmite_2013}) displays the most connections with other proxy records during the period 6,000--3,000 yrs BP. This coincides with a period of dense coupling across the full network. Inspection of the raw data (Figure~\ref{fig:IASMproxies}a) shows this period to be transitional, with a step change in $\delta^{18}O$ values at around 4,000 years BP. This would be expected with a shift to Southern Hemisphere precessional bias, and thus the paleoclimate network seems to capture this regional response to changes in Milankovich forcing.

From 5,000--2,000 years BP onwards, the number of coupling relationships between the IASM proxies and the ITCZ-related and Chinese proxy records begins to decline. Widespread network connectivity also decreases, and the absence of coupling between F07 (Qunf Cave, Oman; \citealp{fleitmann_holocene_2007}) and M14 (Lonar Lake, India; \citealp{menzel_linking_2014}) may indicate that the networks have captured the southward progression of the ITCZ. These two coincident trends suggest that global-scale synchronicity declines in the later Holocene because of decreasing direct insolation in the Northern Hemisphere. This dampens the strength of the Indian and East Asian monsoon, thus weakening interhemispheric coupling relationships with the IASM region. As such, a model emerges whereby Milankovich insolation forcing acts as a control on regional monsoon strength not only through direct radiative forcing, but also indirectly, by modulating the strength of coupling relationships between regional monsoon systems.

This model of indirect insolation control agrees with proxy records from the Indonesian-Australian monsoon region. \cite{denniston_stalagmite_2013} note that although precession and tilt favour the Southern Hemisphere following 6,000 years BP, precipitation over northwest Australia is observed to decline. \cite{liu_coupled_2003} present sea surface temperature feedbacks as an explanatory mechanism for the counterintuitive response to precessional forcing over northwest Australia, while \cite{wyrwoll_sensitivity_2007} demonstrate that tilt as well as precession plays a critical role in determining monsoon precipitation over northwest Australia. Using a coupled ocean-atmospheric model they found that high tilt lead to enhanced monsoon precipitation, even under a Northern Hemisphere precession bias. In fact, the simulation results display changes in interhemispheric outflows to the Southern Hemisphere between different precession and tilt scenarios \citep{wyrwoll_sensitivity_2007}. The observed weakening in coupling relationships between the IASM region and the Northern Hemisphere into the later Holocene suggests that the transient nature of teleconnections between regional climate systems may have played a critical role in determining the response of the Indonesian-Australian monsoon to Milankovich insolation forcing.

By the latest Holocene (3,000--0 years BP) the Laguna Pallacocha record in Peru (M02; \citealp{moy_variability_2002}) is the only record coupled to both the IASM proxy time series. Because this time series is not suitably represented by Brownian motion (unlike, say, the $\delta^{18}$O speleothem time series), we have defined statistical significance based on simulations of a surrogate Poisson process. This Poisson process is derived from an `event' time series created from 90th percentile events within the M02 record, and is well approximated by a Poisson process. Mutual information, as a nonlinear measure of similarity between time series, should be able to quantify the coupling relationships of the Laguna Pallacocha red intensity index (M02; \citealp{moy_variability_2002}). However, the mutual information estimates calculated from these surrogate Poisson time series will not correspond directly with estimates calculated using the original M02 record. Replicating this analysis using a bootstrapping approach to simulate the M02 record returned the same network structures, indicating that this method is suitable. However, we still proceed with caution when drawing conclusions regarding coupling relationships with the Laguna Pallacocha record. In the latest Holocene, dense coupling with the M02 record is observed, including both the IASM records. \cite{moy_variability_2002} interpret the events recorded in the red intensity index as indicative of moderate-to-strong El Ni\~no events. They therefore argue that the record demonstrates an increase in these events from around 3,000 years BP until 1,200 years BP. Other marine and terrestrial proxy records from the Pacific region indicate similar trends -- lake sediment records from Galapagos \citep{conroy_holocene_2008-1}, and proxy records of upwelling on the Peruvian margin \citep{rein_nino_2005} and in the Panama basin \citep{cabarcos_high-resolution_2014}. However, at sub-millennial timescales, the timing of ``spikes" in ENSO activity does not match between records, and there exists a further set of marine records which suggest slight different scenarios -- A sediment core from the Peruvian margin suggests an increase in both El Ni\~no and La Ni\~na strength from 3,000 years BP onwards \citep{makou_postglacial_2010}, and \cite{koutavas_mid-holocene_2006} argue that although the late Holocene experienced an increase in El Ni\~no activity, El Ni\~no strength was not abnormally high, but simply increasing after a period of La Ni\~na-like conditions in the early to mid-Holocene. In the Australian summer monsoon context, a number of authors have suggested that enhanced ENSO activity in the latest Holocene may have contributed to a period of aridity evidenced in speleothem and paleoenvironmental records from northwest Australia \citep{shulmeister_pollen_1995, mcgowan_evidence_2012, denniston_stalagmite_2013}. However, it has been recognised for a long time that there is no apparent impact of ENSO on monsoon precipitation over northwest Australia \citep{mcbride_seasonal_1983}. This is also evident in precipitation records, with the exception of the 1982--1983 El Ni\~no event, during which the Southern Oscillation Index reached a record low of -33 \citep{bureau_of_meteorology_australian_2014}. During this event, there was a clear reduction in Australian summer monsoon rainfall. There is therefore the possibility of an ENSO--northwest Australian teleconnection, although only under an anomalously large shift in the Walker circulation. The paleoclimate networks provided here suggest, albeit tentatively, that a teleconnection between ENSO events and IASM precipitation may well have been present during the latest Holocene. However, given that the Laguna Pallacocha record is, itself, dependent on teleconnections with the central Ni\~no 3.4 region, we make no firmer claims, as the relationship between the Peruvian  and ENSO variability may also be transient during the later Holocene.

In summary, by using the verified complex networks method to identify significant coupling relationships between paleoclimate proxy records, we are able to make the following comments:
\begin{enumerate}
\item The observed coupling relationships between proxy records from the Indian, East Asian and Indonesian Australian monsoon regions highlight the validity of the `global monsoon' model.
\item However, these coupling relationships are observed to change over the last 9,000 years. In particular, a global-scale transition is observed, whereby the paleoclimate network decreases in connectivity towards the later Holocene. This may indicate reduced global-scale coupling caused by Southern Hemisphere precessional bias, causing a southward shift in the ITCZ.
\item In the Australian summer monsoon context, connections are observed with a number of Northern Hemispheric proxy records in the early to mid-Holocene. This corresponds with an understanding of interhemispheric flows modulating monsoon strength over northwest Australia (e.g. \citealp{liu_coupled_2003}). In the latest Holocene, these coupling relationships are lost and the Laguna Pallacocha record (M02; \citealp{moy_variability_2002}) is connected instead. This raises the possibility of an ENSO teleconnection to northwest Australia during this period. This shows some alignment with the dual-model of tropical climate set out by \cite{chiang_tropics_2009}, with global scale controls (i.e. ITCZ) and those derived from regional differences (i.e. ENSO) both playing a role in determining regional monsoon strength throughout the last 9,000 years.
\end{enumerate}

\section{Conclusions}
Identifying potential coupling relationships between climate systems using paleoclimate proxy records is typically a qualitative process. Here, we demonstrate the efficacy of complex networks to identify coupling relationships and teleconnections which correspond to known dynamical mechanisms. Extending this method to a multi-proxy database of paleoclimate time series, we are able to draw conclusions regarding the nature of coupling across the Asian-Australasian monsoon region. Our results recognise an element of the global monsoon concept, with regional monsoons displaying some degree of coupling over the period. The global paleomonsoon model, however, does not adequately represent the transient nature of coupling relationships, while our findings demonstrate a strengthening of coupling relationships across the broad Asian-Australasian monsoon regions during the mid-Holocene, followed by a tendency to reduced coupling in the later Holocene. Our findings at this stage are preliminary, and dependent on the availability of suitable proxy datasets, but we envisage that once more datasets become available, a stronger case can be made. In the context of the Australian summer monsoon, we observe coupling relationships to other low latitude regions throughout the Holocene. While we offer tentative explanations for these, we stress that the observed links are unable to tell us about the underlying mechanism. We can, however, state with confidence that the networks demonstrate effectively that coupling relationships between the northwest Australian monsoon region and other regional climate systems were transient over the past 9,000 years. Given the coupling relationships observed, the next step is to ask why they exist. We recommend further research in the context of available model simulations, and in developing methods to assess the direction of information flow, in an effort to identify the underlying mechanisms determining and driving the coupling relationships.

\section{Acknowledgements}
KHW wishes to thank the Kimberley Foundation Australia for their funding and support. FHM is funded through an Australian Postgraduate Award. Additionally, the authors wish to thank Jenny Hopwood for valuable discussions.

\bibliography{transientcouplingrelationships_revisedARXIV}

\end{document}